# An ordinary differential equation for velocity distribution and dip-phenomenon in open channel flows

Rafik ABSI,
Assoc. Professor*, EBI, Inst. Polytech. St-Louis, Cergy University, 32 boulevard du Port, 95094 Cergy-Pontoise cedex, France. E-mail: r.absi@ebi-edu.com ; rafik.absi@yahoo.fr*

## ABSTRACT

An ordinary differential equation for velocity distribution in open channel flows is presented based on an analysis of the Reynolds-Averaged Navier-Stokes equations and a log-wake modified eddy viscosity distribution. This proposed equation allows to predict the velocity-dip-phenomenon, i.e. the maximum velocity below the free surface. Two different degrees of approximations are presented, a semi-analytical solution of the proposed ordinary differential equation, i.e. the full dip-modified-log-wake law and a simple dip-modified-log-wake law. Velocity profiles of the two laws and the numerical solution of the ordinary differential equation are compared with experimental data. This study shows that the dip correction is not efficient for a small Coles' parameter, accurate predictions require larger values. The simple dip-modified-log-wake law shows reasonable agreement and seems to be an interesting tool of intermediate accuracy. The full dip-modified-log-wake law, with a parameter for dip-correction obtained from an estimation of dip positions, provides accurate velocity profiles.

## 1 Introduction

Due to practical implications, the velocity distribution of open-channel flows has interested engineers and researchers for many years. The vertical velocity profile is well described by the classical log law in the inner region $\xi < 0.2$, where $\xi = y/h$ is the ratio of the distance from the bed to flow depth (Nezu and Rodi 1986, Cardoso *et al.* 1989, Nezu and Nakagawa 1993, Li *et al.* 1995). However, the log law normally deviates from experimental data in the outer region $\xi > 0.2$. This deviation is accounted for by adding Coles' wake function (Coles 1956, Hinze 1975). In two-dimensional (2D) open-channel flows, in addition to the simple power law (Afzal 2005, Castro-Orgaz 2009), the log-wake law appears to be the most reasonable extension of the log law (Nezu and Nakagawa 1993). However, in narrow open-channels involving an aspect ratio $Ar < 5$, where $Ar = b/h$ is the ratio of the channel width $b$ to flow depth, and near side walls or corner zones even for wide open-channels (Vanoni 1941), the maximum velocity appears below the free surface producing the velocity-dip-phenomenon, involving a deviation from the log-wake law. This phenomenon, which was reported more than a century ago (Francis 1878, Stearns 1883), was observed both in open-channels and rivers. It is related to secondary currents generated in three-dimensional (3D) open-channel flows (Imamoto and Ishigaki 1988, Wang and Cheng 2005). Coles' wake function is unable to represent this behavior since it predicts a velocity increasing with distance from the bed.

The standard two-equation $k$-$\varepsilon$ model is unable to predict secondary currents and the related velocity-dip-phenomenon since it assumes isotropic turbulence. Accurate predictions of velocity-dip-phenomena require therefore more sophisticated Reynolds-Averaged Navier

Stokes (RANS) -based anisotropic turbulence models such as the Reynolds Stress Model (RSM) (Kang and Choi 2006). Large Eddy Simulation (LES) and Direct Numerical Simulation (DNS) allow predicting secondary currents in narrow open-channels (Hayashi *et al.* 2006). Instead of these turbulence models, analytical and/or empirical relations were proposed to predict the velocity-dip phenomenon for engineering applications. Sarma *et al.* (2002) proposed a generalized version of the binary velocity distribution law, which combines the logarithmic law of the inner region with the parabolic law for the outer region. Guo and Julien (2003, 2008) proposed a Modified-Log-Wake law (MLW-law) which fits velocity profiles with dip-phenomenon. However, this law cannot be used for predictive applications since it requires to fit the near free surface velocities to the parabolic law to obtain dip position and maximum velocity (Guo and Julien 2008). As they indicate, the MLW-law can be used only in flow measurements since it requires sampled velocities. Yang *et al.* (2004) proposed a Dip-Modified-Log law (DML-law) based on the analysis of the RANS equations. This law, involving two logarithmic distances, one from the bed (i.e. the log law), and the other from the free surface, has the advantage that it contains only parameter $\alpha$ for dip-correction. The DML-law reverts to the classical log law for $\alpha=0$.

In this study, it is first pointed out that even if the dip-modified-log law (Yang *et al.* 2004) predicts dip-phenomena well for smooth uniform open channel flows, it deviates from experimental data in the rough wall flow regime. In most cases, it is impossible to improve the velocity profiles by adjusting parameter $\alpha$. The aim of this study is to improve this prediction for arbitrary open channel flows. Section 2 presents model equations based on the RANS equations, and the related assumptions. Dip-modified laws will be presented in Section 3; a simple Dip-Modified-Log-Wake law (sDMLW-law), which reverts to log-wake law for large values of $Ar$, will be presented. In Section 4, an Ordinary Differential Equation (ODE) for open channel velocity distribution is presented using a log-wake modified eddy viscosity distribution. Numerical and semi-analytical solutions involving the full Dip-Modified-Log-Wake law (fDMLW-law) will be validated by experimental data.

## 2 Model equations

For steady uniform open-channel flows, using the continuity equation, the RANS momentum equation reads in the streamwise direction $x$ (Fig. 1)

$$\frac{\partial U\,V}{\partial y} + \frac{\partial U\,W}{\partial z} = \nu\,\frac{\partial^2 U}{\partial y^2} + \nu\,\frac{\partial^2 U}{\partial z^2} + \frac{\partial -\overline{u\,v}}{\partial y} + \frac{\partial -\overline{u\,w}}{\partial z} + g\,\sin\theta \tag{1}$$

where $x$, $y$ and $z$ are respectively streamwise, vertical and lateral directions and $U$, $V$ and $W$ the three corresponding mean velocities with $u$, $v$ and $w$ as turbulent fluctuations, $\nu$ the fluid kinematic viscosity, $g$ the gravitational acceleration, and $\theta$ is the angle of the channel bed to the horizontal (Fig. 1b). Equation (1) may be written with $S=\sin\theta$ as channel bed slope as

$$\frac{\partial\left(U\,V - \left[\nu\,\dfrac{\partial U}{\partial y} - \overline{u\,v}\right]\right)}{\partial y} + \frac{\partial\left(U\,W - \left[\nu\,\dfrac{\partial U}{\partial z} - \overline{u\,w}\right]\right)}{\partial z} = g\,S \tag{2}$$

In the central channel zone (Fig. 1a), it is assumed that the vertical gradients (d/d$y$) are dominating, allowing to therefore neglect the horizontal gradients (d/d$z$) (Yang *et al.* 2004). Since for large values of $y$ the viscous part ($\nu$·d$U$/d$y$) of the shear stress $\tau/\rho=(\nu\,\mathrm{d}U/\mathrm{d}y)-\overline{u\,v}$, where $\rho$ is fluid density, is small versus the turbulent part $-\overline{u\,v}$ (Absi 2008), Eq. (2) becomes

$$\frac{\partial U\,V}{\partial\,y} + \frac{\partial\,\overline{u\,v}}{\partial\,y} = g\,S \tag{3}$$

Integration of Eq. (3) gives

$$\frac{-\overline{u\,v}}{u_*^2} = \left(1 - \frac{y}{h}\right) - \alpha_1\frac{y}{h} + \frac{U\,V}{u_*^2} \tag{4}$$

where $u_*$ is friction velocity and $\alpha_1 = [(gSh)/u_*^2] - 1$.

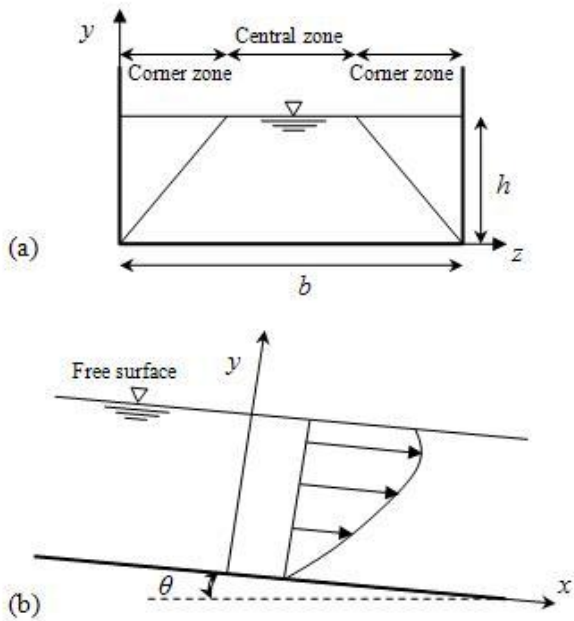

Figure 1 Definition sketch for steady uniform open-channel flow

By assuming (Yang *et al.* 2004)

$$\frac{U\,V}{u_*^2} \approx -\alpha_2\frac{y}{h} \tag{5}$$

where $\alpha_2$ is a positive coefficient, Eq. (4) becomes

$$\frac{-\overline{u\,v}}{u_*^2} = \left(1 - \frac{y}{h}\right) - \alpha\frac{y}{h} \tag{6}$$

where $\alpha = \alpha_1 + \alpha_2$. With the Boussinesq assumption

$$-\overline{u\,v} = \nu_t\frac{\mathrm{d}\,U}{\mathrm{d}\,y} \tag{7}$$

Eq. (6) gives

$$\frac{\mathrm{d}U}{\mathrm{d}y} = \frac{u_*^2}{\nu_t}\left[\left(1 - \frac{y}{h}\right) - \alpha\,\frac{y}{h}\right] \tag{8}$$

Equation (8) contains two unknowns, namely d$U$/d$y$ and $\nu_t$. Since the aim of this study is to predict velocity profiles in open-channel flows, the eddy viscosity $\nu_t$ is required.

### 3 Dip-modified laws
#### 3.1 Dip-modified log law

With a known eddy viscosity profile $\nu_t(y)$, integration of Eq. (8) provides the velocity distribution. Yang *et al.* (2004) obtained a DML-law based on Eq. (8) and a parabolic eddy viscosity

$$\nu_t = \kappa\,u_*\,y\left(1 - \frac{y}{h}\right) \tag{9}$$

where $\kappa \approx 0.41$ is the von Karman constant. Equation (9) allows to express Eq. (8) as

$$\frac{\mathrm{d}U}{\mathrm{d}y} = \frac{u_*}{\kappa\,y}\left(1 - \alpha\,\frac{\dfrac{y}{h}}{1 - \dfrac{y}{h}}\right) \tag{10}$$

Integration of Eq. (10) gives (Yang *et al.* 2004)

$$\frac{U}{u_*} = \frac{1}{\kappa}\left[\ln\left(\frac{y}{y_0}\right) + \alpha\ln\left(\frac{1 - \dfrac{y}{h}}{1 - \dfrac{y_0}{h}}\right)\right] \tag{11}$$

where $y_0$ is the distance at which the velocity is hypothetically equal to zero. Since $y_0/h \ll 1$ and with $U_a = U/u_*$, $\xi = y/h$ and $\xi_0 = y_0/h$, Eq. (11) simplifies to

$$U_a = \frac{1}{\kappa}\left[\ln\left(\frac{\xi}{\xi_0}\right) + \alpha\ln\left(1 - \xi\right)\right] \tag{12}$$

The dip-modified log law predicts the velocity-dip-phenomenon by the term $\ln(1 - y/h)$ of Eq. (12), and $\alpha$ as dip-correction parameter (Yang *et al.* 2004). This law contains only $\alpha$ and reverts to the classical log law for $\alpha = 0$.

Yang *et al.* (2004) proposed the empirical formula $\alpha(z) = 1.3\exp(-z/h)$, where $z$ is the lateral distance from the side wall. On the channel axis at $z = b/2$ and therefore $z/h = Ar/2$, this equation reads $\alpha(Ar) = 1.3\exp(-Ar/2)$. A single general formula may be expressed as

$$\alpha = C_1\exp\left(-C_2\,Ar\,Z\right) \tag{13}$$

where $Z = z/(b/2) = (2z)/b$ is dimensionless lateral distance from the side wall. From calibration Yang *et al.* (2004) found for the two coefficients $C_1 = 1.3$ and $C_2 = 0.5$. For wide open-channels ($Ar > 5$), $\alpha \rightarrow 0$ (Eq. 13) and the DML-law (Eq. 12) reverts to the log-law. However, the log-law is valid only in the inner region ($\xi = y/h < 0.2$).

### 3.2 *Simple dip-modified-log-wake law*

In the outer region ($\xi > 0.2$), the log-law deviates from experimental data. In 2D open-channel flows, this deviation is accounted for by adding Coles (1956) wake function $(2\Pi/\kappa)\cdot\sin^2(\pi y/2h)$ as (Hinze 1975)

$$\frac{U}{u_*} = \frac{1}{\kappa}\left[\ln\left(\frac{y}{y_0}\right) + 2\,\Pi\sin^2\left(\frac{\pi y}{2h}\right)\right]$$

(14)

where $\Pi$ is Coles' parameter expressing the strength of the wake function. This log-wake law appears to be the most reasonable extension of the log-law. However, the value of $\Pi$ seems to be not universal. Cebeci and Smith (1974) found experimentally that $\Pi$ increases with the Reynolds number in zero-pressure-gradient boundary layers, attaining an asymptotic value of $\Pi=0.55$ at high Reynolds numbers. Laser Doppler Anemometry (LDA) velocity measurements in 2D fully-developed open-channel flow over smooth beds (Nezu and Rodi 1986) indicated that $\Pi$ increases from zero with the friction Reynolds number $\mathsf{R}_*=u_*h/\nu$ and becomes nearly constant $\Pi\approx0.2$ for $\mathsf{R}_*>2,000$ or $\mathsf{R}_h=4hU_m/\nu>10^5$, where $U_m$ is the mean bulk velocity. Cardoso *et al.* (1989) observed for uniform flow in a smooth open channel in the core of the outer region ($0.2<y/h<0.7$) a wake of a relatively small strength ($\Pi\approx0.08$), followed in the near-surface-zone ($0.7<y/h<1$) by a retarding flow associated with weak secondary currents. Velocity measurements by Kirkgoz (1989), in fully-developed rectangular subcritical open channel flow on smooth bed, indicate $\Pi=0.1$, whereas measurements of Li *et al.* (1995) give $\Pi=0.3$ for Froude numbers $\mathsf{F}=U_{max}/(gh)^{1/2}>1$ and $\mathsf{R}_h>10^5$, where $\mathsf{R}_h$ and $\mathsf{F}$ are defined by the maximum velocity $U_{max}$ instead of $U_m$.

In 3D open-channel flows with secondary currents, the log-wake law is unable to predict the velocity-dip-phenomenon. A suitable simplification results by adding both Eq. (14) and the term $\ln(1-y/h)$ of Eq. (12) to the log law as (Absi 2009)

$$U_a = \underbrace{\frac{1}{\kappa}\ln\left(\frac{\xi}{\xi_0}\right)}_{\text{Term I}} + \underbrace{\frac{2\Pi}{\kappa}\sin^2\left(\frac{\pi}{2}\xi\right)}_{\text{Term II}} + \underbrace{\frac{\alpha}{\kappa}\ln\left(1-\xi\right)}_{\text{Term III}}$$

(15)

Equation (15) is referred as the simple dip-modified-log-wake law (sDMLW-law) of which the advantage is that it reverts to the log-wake law ($\alpha=0$) in 2D open-channel flows.

## 4 Ordinary differential equation for velocity distribution and dip phenomenon
### 4.1 *Ordinary differential equation*

Instead of the parabolic profile for eddy viscosity (Eq. 9), a more appropriate approximation in accordance with the log-wake law given by Nezu and Rodi (1986) is

$$\frac{\nu_t}{u_*h} = \kappa\left(1-\frac{y}{h}\right)\left[\frac{h}{y} + \pi\,\Pi\sin\left(\frac{\pi y}{h}\right)\right]^{-1}$$

(16)

Inserting Eq. (16) into Eq. (8), the ODE for velocity distribution reads

$$\frac{\mathrm{d}U}{\mathrm{d}y} = \frac{u_*}{\kappa\, h}\left(1 - \alpha\,\frac{\dfrac{y}{h}}{1 - \dfrac{y}{h}}\right)\left[\frac{h}{y} + \pi\,\Pi\sin\left(\frac{\pi\, y}{h}\right)\right]$$

(17)

For $\alpha=0$, Eq. (17) gives the log-wake law. Equation (17) may also be written as

$$\frac{\mathrm{d}U}{\mathrm{d}y} = \frac{u_*}{\kappa\, y}\left(1 - \alpha\,\frac{\dfrac{y}{h}}{1 - \dfrac{y}{h}}\right)\left[1 + \pi\,\Pi\,\frac{y}{h}\sin\left(\frac{\pi\, y}{h}\right)\right]$$

(18)

For $\Pi=0$, Eq. (18) reverts to Eq. (10) providing Eq. (12). In dimensionless form Eq. (18) reads

$$\frac{\mathrm{d}U_a}{\mathrm{d}\xi} = \frac{1}{\kappa}\left(1 - \alpha\,\frac{\xi}{1-\xi}\right)\left[\frac{1}{\xi} + \pi\,\Pi\sin\left(\pi\,\xi\right)\right]$$

(19)

For $\alpha=0$ and $\Pi=0$, integration of Eq. (19) gives the log law, ~~providing~~. Eq. (19) provides the *dip* (subscript *dip*) distance $\xi_{dip}$ or the dimensionless distance from the bed corresponding to the maximum velocity $U_{a,max}=U_a(\xi=\xi_{dip})$. Since $\mathrm{d}U_a/\mathrm{d}\xi=0$ at $\xi=\xi_{dip}$, $\alpha$ is given from Eq. (19) by

$$\alpha = \frac{1}{\xi_{dip}} - 1$$

(20)

The maximum velocity occurs therefore at $\xi_{dip}=1/(\alpha+1)$. Eq. (6) shows that the Reynolds shear stress is equal to zero at $\xi_{dip}$. This seems indicate that zero Reynolds stress corresponds to maximum velocity. The numerical solution of Eq. (19) is found by using the 4[th]-order Runge-Kutta scheme or by the "ode45" function of MATLAB (MATLAB 2003).

### 4.2 *Full dip-modified-log-wake law*
Integration of Eq. (19) for $\xi_0 \ll 1$ gives

$$U_a = \underbrace{\frac{1}{\kappa}\ln\left(\frac{\xi}{\xi_0}\right)}_{\text{Term I}} + \underbrace{\frac{2\,\Pi}{\kappa}\sin^2\left(\frac{\pi}{2}\,\xi\right)}_{\text{Term II}} + \underbrace{\frac{\alpha}{\kappa}\ln(1-\xi)}_{\text{Term III}} - \underbrace{\frac{\alpha\,\pi\,\Pi}{\kappa}\int_{\xi_0}^{\xi}\frac{\xi}{1-\xi}\sin\left(\pi\,\xi\right)\mathrm{d}\xi}_{\text{additional term IV}}$$

$$\underbrace{\phantom{XX}}_{\text{sDMLW-law}}$$

(21)

Equation (21) is referred as the full dip-modified-log-wake law (fDMLW-law). This equation differs from the simple dip-modified-log-wake law (sDMLW-law, Eq.15) only by the additional term IV. To evaluate the difference between Eq. (21) and Eq. (15), term IV needs to be integrated using ~~the~~ trapezoidal or Simpson rules. For wide open-channels (*Ar*>0.5), $\alpha \rightarrow 0$, and the fDMLW-law reverts to log-wake law since terms III and IV then vanish.

## 5 Results and discussion

Figure 2 compares the predicted velocity profiles by log-law, dip-modified-log law (DML-law, Eq. 12) and log-wake-law (Eq. 14) with experimental data of Coleman (1986), Sarma *et al.* (2000) and Lyn (1986). Plots 2(a.1) and 2(c.1) show velocity profiles at the channel axis, while velocity profiles of 2(b.1) are at a certain distance from the lateral side wall. This figure shows that the log-law is able to predict experimental data in the inner region. The DML-law with $\alpha$ from (Eq.13) allows to predict a deviation from the log-law with a maximum velocity below the free surface (i.e. the velocity-dip-phenomenon). However, the DML-law profiles differ from the experimental data. In plots 2(a.1) and 2(c.1), it is impossible to improve the predicted velocity profiles from the DML-la by adjusting parameter $\alpha$, since the deviation increases by increasing $\alpha$ and the predicted profiles approach the log-law profiles otherwise. In Fig. 2(b.1), it is possible to improve the DML-law velocity profile by decreasing $\alpha$ via the parameters $C_1$ and $C_2$ in Eq. (13). Log-wake-law profiles (2(a.1) and 2(c.1)) with parameter $\Pi=0.2$ allow to improve predicted velocities up to $\xi=0.4$. However, the value of $\Pi=0.2$ seems to be irrelevant for an accurate dip correction. With $\Pi=0.2$, the dip-modified-log-wake laws are able to predict the velocity-dip-phenomenon but with inaccurate maximum velocities. Accurate predictions of the velocity-dip-phenomenon require therefore larger values of $\Pi$. Figure 2(a.2), (b.2) and (c.2) shows velocity profiles in defect form.

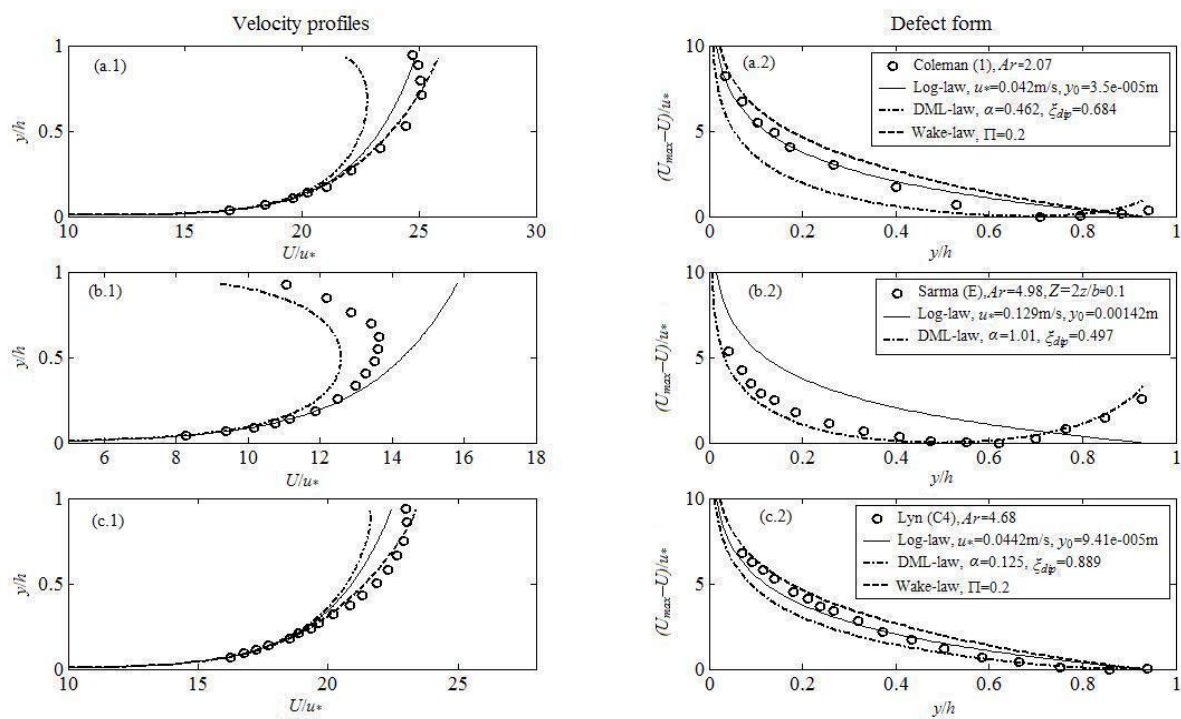

Figure 2 Velocity profiles, comparison between log-law, dip-modified-log law (DML-law, Eq. 12) with $\alpha$ from Eq. (13), log-wake-law (Eq. 14) with $\Pi=0.2$, and experimental data of Coleman (1986) Run 1, Sarma *et al.* (2000) Run E and Lyn (1986) Run C4; (a.2), (b.2) and (c.2) are in defect form

Figure 3 compares velocity profiles obtained from the log-law, log-wake-law (Eq. 14), simple dip-modified-log-wake law (sDMLW-law, Eq. 15) and the full dip-modified-log-wake law (fDMLW-law, Eq. 21) with experimental data of Coleman (1986), Sarma *et al.* (2000) and Lyn (1986). Term II of the wake function in the sDMLW-law associated with term III of the DML-law is important for an adequate prediction of the dip-phenomenon. Since $\Pi=0.2$ seems to be too small to improve predictions, $\Pi = 0.45$ was used, while $\alpha$ was obtained by Eq. (13) calibrated by Yang *et al.* (2004). Integration in term IV is resolved by the "quad" function of MATLAB (MATLAB 2003) which uses recursive adaptive Simpson quadrature. Plots 3(a.2), (b.2) and (c.2) show the effect of each term in the sDMLW-law (Eq. 15) and the fDMLW-law (Eq. 21). Term II increases velocities while terms III and IV have a decreasing effect. Velocity profiles obtained by Eq. (15) with $\alpha$ from Eq. (13) and calibrated by Yang *et al.* (2004) show reasonable agreement. Term IV improves velocity profiles only in Fig. 3(c.1). In Fig. 3(a.1) and (b.1), the sDMLW-law is more accurate whereas the fDMLW-law deviates from experimental data, which seems to be related to $\alpha$ obtained by Eq. (13) and does not correspond to measured dip positions (Eq. 20).

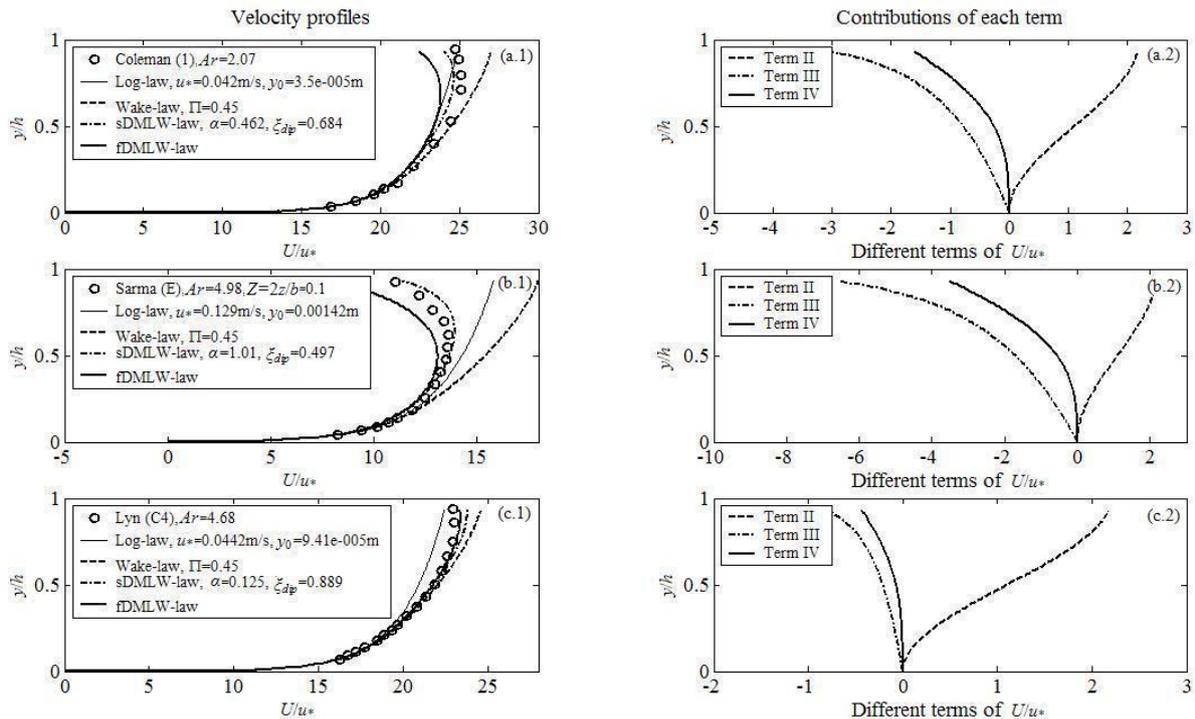

Figure 3 Velocity profiles with $\Pi = 0.45$ and $\alpha$ from Eq. (13). Comparison between log-law, log-wake-law (Eq. 14), sDMLW-law (Eq. 15), fDMLW-law (Eq. 21) and experimental data. (a.2), (b.2) and (c.2) show effect of each term in sDMLW-law and fDMLW-law

In Figure 4, values of $\alpha$ obtained from Eq. (20) were used with $\Pi = 0.45$. In addition to the log-law, the log-wake-law (Eq. 14), the sDMLW-law (Eq. 15) and the fDMLW-law (Eq. 21), Fig. 4(a.1), (b.1) and (c.1) show numerical solutions of Eq. (19) obtained by the "ode45" function of MATLAB. The fDMLW-law and numerical solutions provide accurate velocity profiles. Figure 4 shows no differences between the fDMLW-law profiles and the numerical solutions of Eq. (19). These comparisons demonstrate the accuracy of the fDMLW-law.

Figure 5 compares predicted velocity profiles with Runs C1 to C4 of Lyn (1986) in semi-log scales to show the deviation from the log-law. Figure 5(a) and (b) are for $Ar = 4.1$ and $S = 2.06 \times 10^{-3}$ and $S = 2.7 \times 10^{-3}$, while Fig. 5(c) and (d) relates to $Ar = 4.68$ and $S = 2.06 \times 10^{-3}$ and $S = 4.01 \times 10^{-3}$ for $\Pi = 0.45$ with $\alpha$ obtained by Eq. (13) for the DML-law and by estimating the dip positions $\xi_{dip}$ from experimental data for sDMLW-law and fDMLW-law. The fDMLW-law profiles agree well with the experimental data. For the same aspect ratio, velocity profiles require two different values of $\alpha$. This suggests that $\alpha$ should depend also on channel slope, in addition to the aspect ratio and the lateral distance from the side wall.

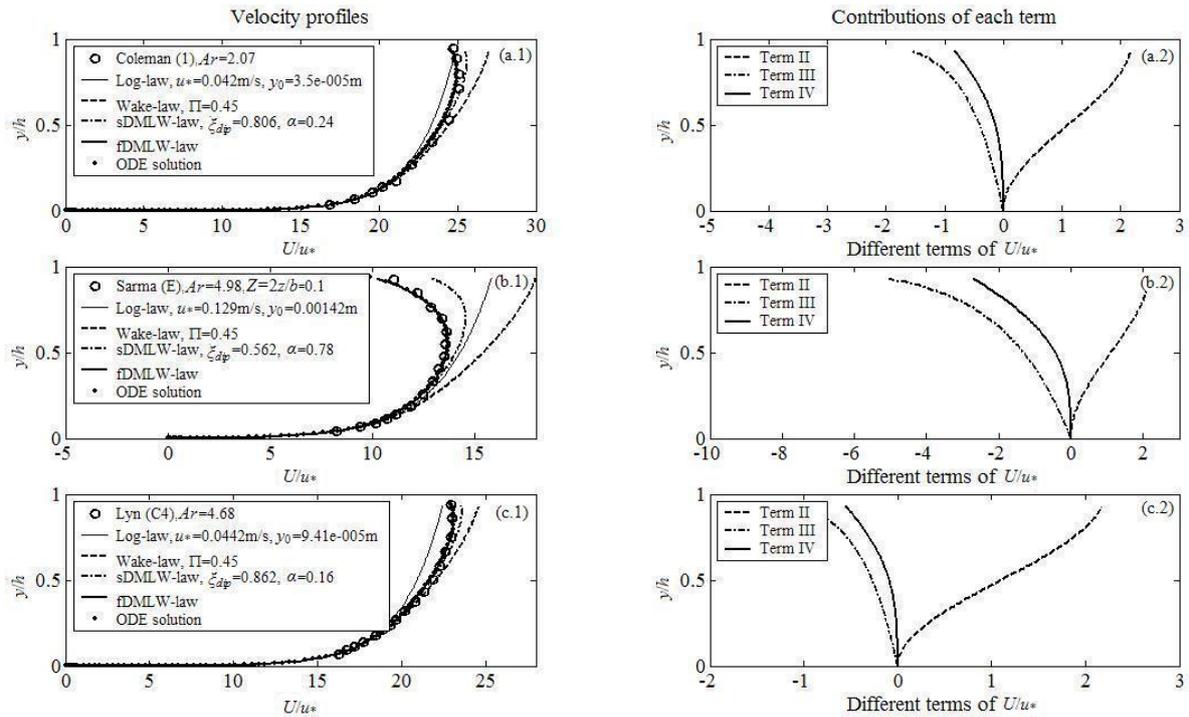

Figure 4 Velocity profiles with $\Pi=0.45$ and $\alpha$ from Eq. (20). Comparison between log-law, log-wake-law (Eq. 14), sDMLW-law, (Eq. 15), fDMLW-law (Eq. 21), numerical solutions of Eq. (19) and experimental data. Plots (a.2), (b.2) and (c.2) show effect of each term in sDMLW-law and fDMLW-law

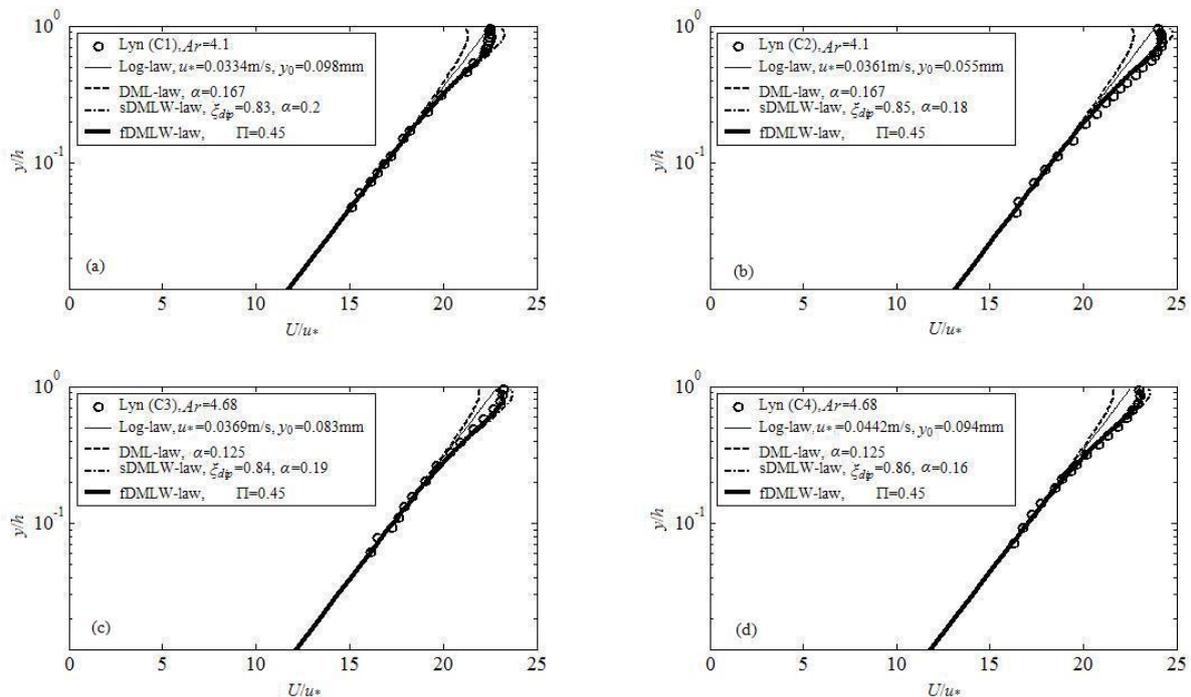

Figure 5 Predicted velocity profiles in semi-log scales with experimental data of Lyn (1986), Runs C1, C2, C3 and C4

## 6 Conclusions

An ordinary differential equation for the velocity distribution and the dip phenomenon in open channel flows is proposed based on an analysis of the Reynolds-averaged Navier-Stokes equations. It was obtained using a log-wake modified eddy viscosity distribution. A semi-analytical solution resulting in the full dip-modified-log-wake law was obtained. Two different degrees of approximation are presented, a more accurate full dip-modified-log-wake law and a simple dip-modified-log-wake law. The first differs from the second law only by an additional term which requires integration. Velocity profiles of the two laws, where the integral in the additional term was resolved ~~by the trapezoidal rule for numerical integration~~ by the "quad" function of MATLAB which uses recursive adaptive Simpson quadrature, and the numerical solution of the ODE obtained by the "ode45" function of MATLAB, were compared to experimental data. The contribution of each term in the two laws was evaluated. The dip correction is not important if $\Pi$ remains small. Accurate predictions of the velocity-dip-phenomenon require larger values. Velocity profiles obtained by the simple dip-modified-log-wake law with $\Pi$=0.45 and calibrated $\alpha$ show reasonable agreement. This law, with $\alpha$ given by Eq. (13), seems to be an interesting tool of intermediate accuracy. The full dip-modified-log-wake law with $\Pi$=0.45 and $\alpha$ obtained from an estimation of dip positions, provides accurate velocity profiles. However, a more accurate formula and/or calibration of parameter $\alpha$ is needed. Results suggest that parameter $\alpha$, which depends on the aspect ratio and the lateral distance from the side wall, should depend also on the channel slope. The proposed equation and its semi-analytical solution require a deeper analysis.

**Acknowledgements**

The author would like to thank Dr. Junke Guo, University of Nebraska, Lincoln, for having provided the experimental data files.

**Notation**

$Ar = b/h$ = aspect ratio
$b$ = channel width
$C_1$, $C_2$ = coefficients of parameter $\alpha$
$\mathsf{F} = U/(gh)^{1/2}$ = Froude number
$g$ = gravitational acceleration
$h$ = flow depth
$\mathsf{R}_h = 4hU/\nu$ = Reynolds number
$\mathsf{R}_* = u_* h/\nu$ = friction Reynolds number
$S$ = channel slope $\sin\theta$
$U$, $V$, $W$ = respectively streamwise, vertical and lateral mean velocities
$U_a$ = dimensionless streamwise mean velocity
$U_m$ = mean bulk velocity
$U_{max}$ = maximum velocity
$u$, $v$, $w$ = respectively streamwise, vertical and lateral turbulent velocity fluctuations
$u_*$ = friction velocity
$x$, $z$ = streamwise and lateral directions, respectively
$y$ = vertical distance from bed
$y_0$ = distance at which velocity is hypothetically equal to zero
$Z = z/(b/2)=(2z)/b$ = dimensionless lateral distance from side wall
$\alpha$ = parameter related to $Ar$, $Z$ and $\xi_{dip}$
$\alpha_1$, $\alpha_2$ = coefficients
$\Pi$ = Coles' parameter
$\nu$ = kinematic fluid viscosity
$\rho$ = fluid density
$\tau$ = shear stresses
$\theta$ = angle of channel bed to horizontal axis

$\xi = y/h$ = dimensionless distance from bed
$\xi_{dip}$ = dip distance of maximum velocity $U_{a,max} = U_a(\xi = \xi_{dip})$